\documentclass[12pt]{article}
\usepackage{graphicx}

\newcommand\pubnumber{SNSN-323-63}
\newcommand\pubdate{\today}

\def\institute{on behalf of the ATLAS and CMS Collaborations\\Vrije Universiteit Brussel, Brussels, Belgium}
\def\support{\footnote{supported by Research Foundation Flanders (FWO-Vlaanderen).}}

\def\Title#1{\begin{center} {\Large #1 } \end{center}}
\def\Author#1{\begin{center}{ \sc #1} \end{center}}
\def\Address#1{\begin{center}{ \it #1} \end{center}}

\newcommand\pubblock{\rightline{\begin{tabular}{l} \pubnumber\\
         \pubdate  \end{tabular}}}
\newenvironment{Abstract}{\begin{quotation}  }{\end{quotation}}
\newenvironment{Presented}{\begin{quotation} \begin{center} 
             PRESENTED AT\end{center}\bigskip 
      \begin{center}\begin{large}}{\end{large}\end{center} \end{quotation}}

\def\beq{\begin{equation}}
\def\eeq#1{\label{#1}\end{equation}}
\def\eeqn{\end{equation}}

\def\beqa{\begin{eqnarray}}
\def\eeqa#1{\label{#1}\end{eqnarray}}
\def\eeqan{\end{eqnarray}}

\let\bar=\overbar

\def\Dslash{\not{\hbox{\kern-4pt $D$}}}
\def\dslash{\not{\hbox{\kern-2pt $\del$}}}

\def\msb{{\bar{\ssstyle M \kern -1pt S}}}

\begin{document}
\begin{titlepage}
\pubblock

\vfill
\Title{Top quark properties}
\vfill
\Author{Petra Van Mulders\support}
\Address{\institute}
\vfill
\begin{Abstract}
The multi-purpose experiments at CERN's Large Hadron Collider have a very rich programme in top quark physics. The large amount of data allows for measuring the top quark properties with an unprecedented precision. This document presents some of the properties that have been measured using top quark pair events produced in proton-proton collisions with a centre-of-mass energy of 13 TeV. The focus lies on the measurements of the colour flow, the charge asymmetry and spin correlations in top quark pair events.
\end{Abstract}
\vfill
\begin{Presented}
$11^\mathrm{th}$ International Workshop on Top Quark Physics\\
Bad Neuenahr, Germany, September 16--21, 2018
\end{Presented}
\vfill
\end{titlepage}
\def\thefootnote{\fnsymbol{footnote}}
\setcounter{footnote}{0}


\section{Introduction}

The top quark properties can be divided into different types. There are intrinsic properties such as the mass, which is the topic of a dedicated proceeding. Other properties are either related to the top quark decay, e.g. the $W$ boson helicity, or to the production mechanism, e.g. the charge asymmetry or spin correlation in pair-produced top quark events. A last type of ``properties'' is related to the modelling of top quark events in a specific collider environment, e.g. the colour flow. While the presentation at the workshop included more material, this document focuses on the measurements which have been performed using the 13~TeV proton collision data.

\section{Colour flow in top quark pair events}
\begin{figure}[htbp]
\centering
\includegraphics[width=0.32\textwidth]{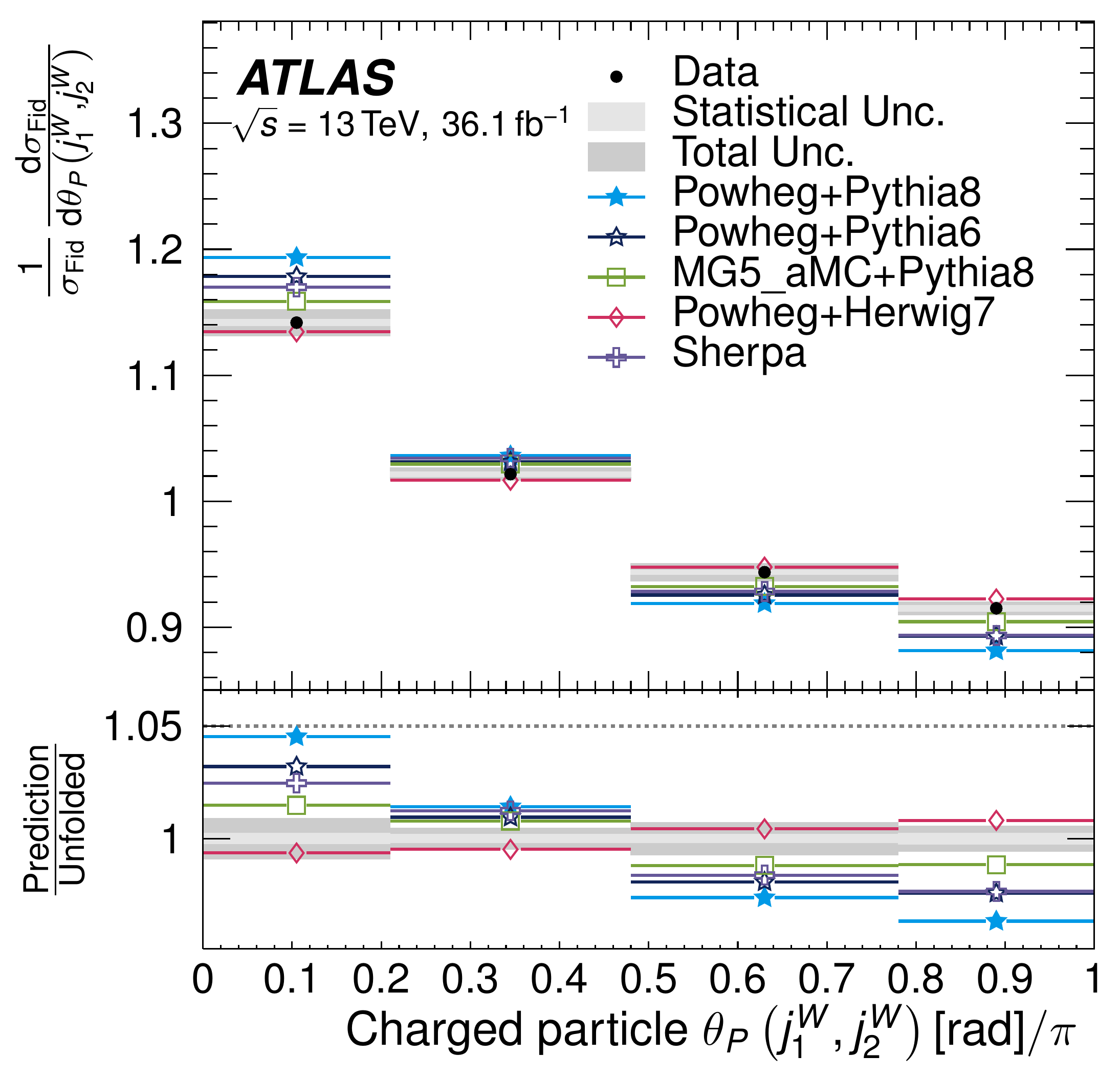}
\includegraphics[width=0.32\textwidth]{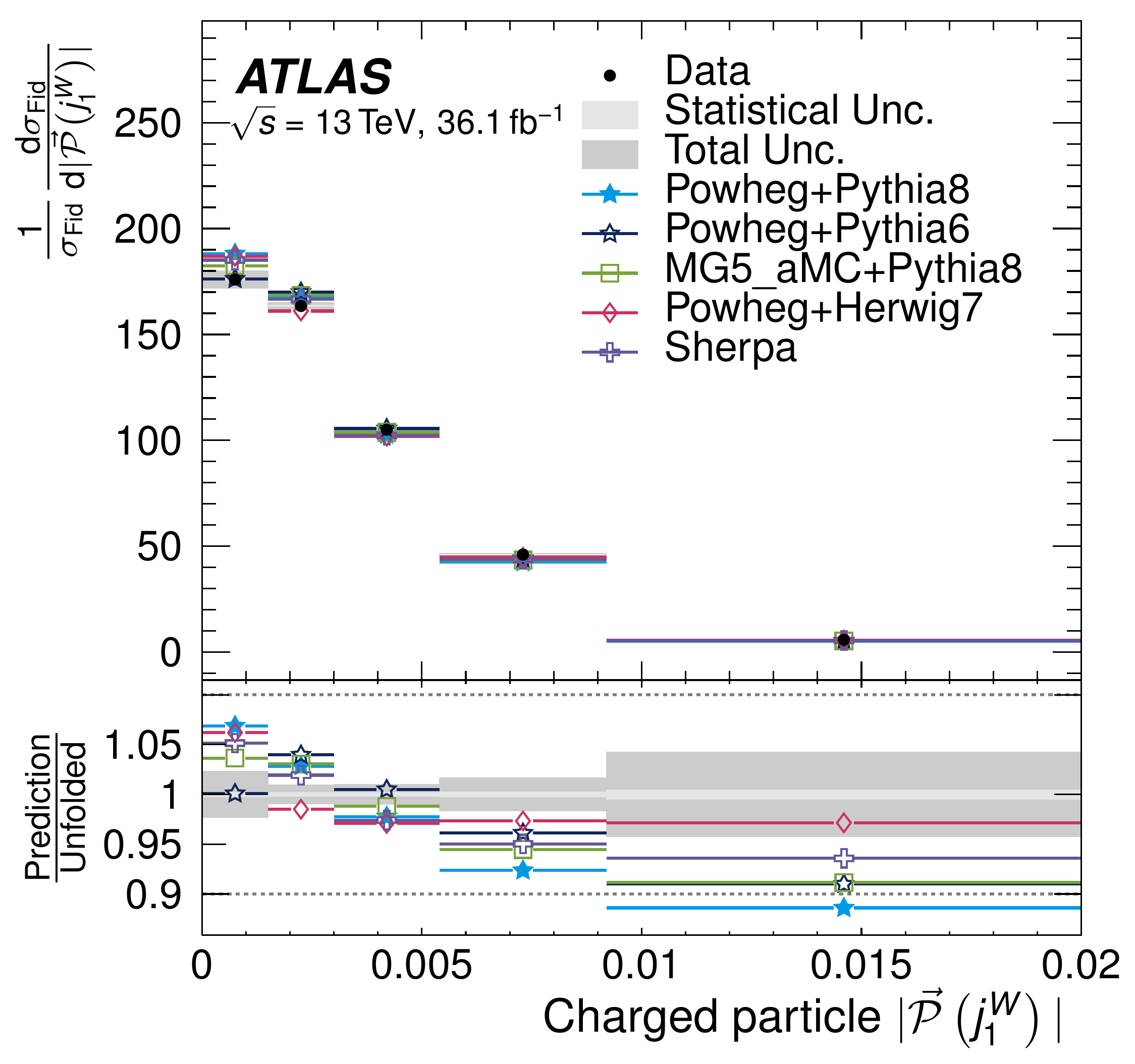}
\includegraphics[width=0.32\textwidth]{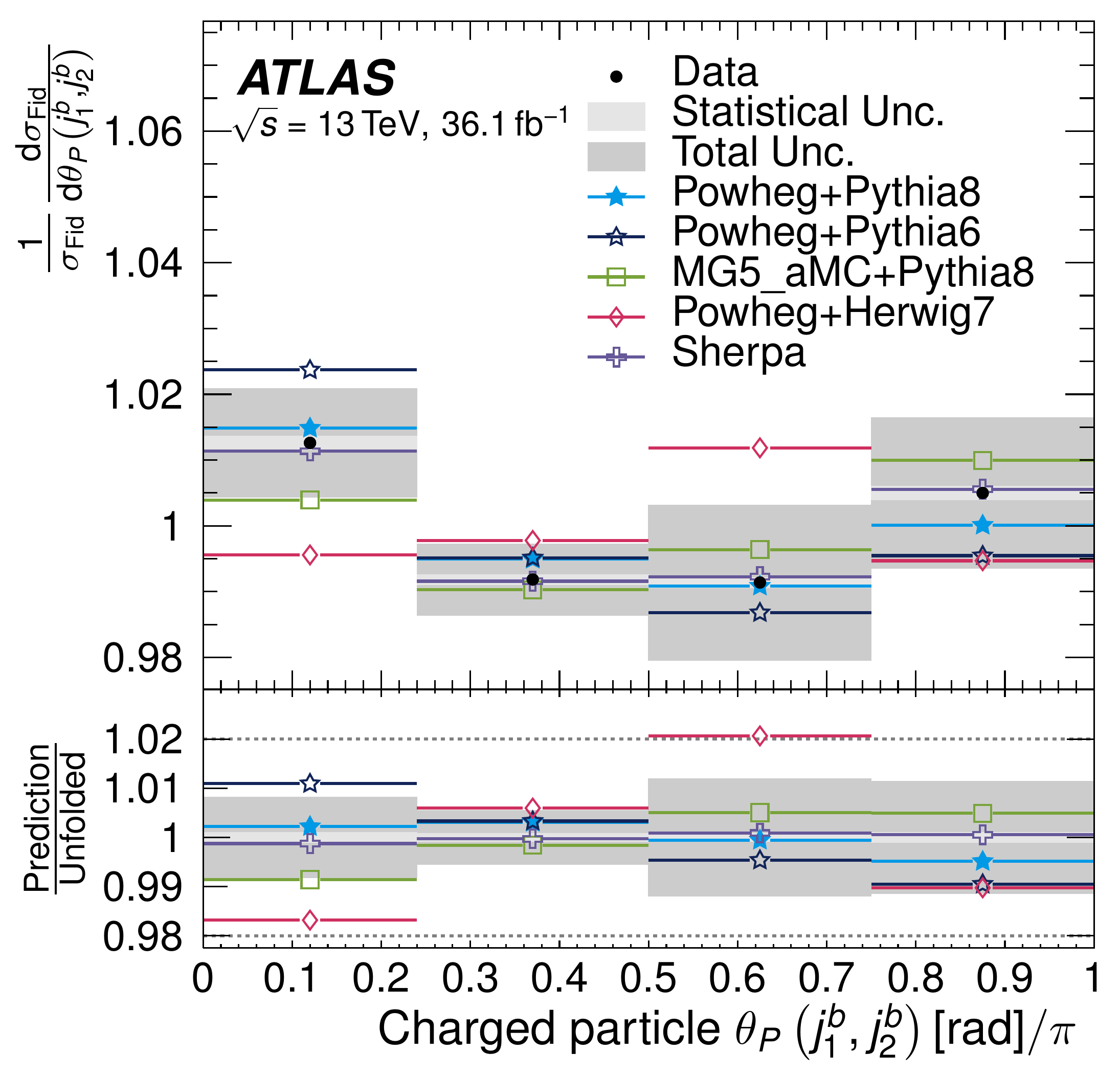}
\caption{The jet-pull angle between the two jets from the hadronically decaying $W$ boson (left), the magnitude of the jet-pull vector for the leading $p_{\mathrm{T}}$ jet from the hadronically decaying $W$ boson (middle), and the jet-pull angle between the two jets originating from the $b$ quarks in the $t\bar{t}$ decay (right)~\cite{ATLAScolourFlow}.}
\label{fig:colourflow}
\end{figure}

Because colour charge is conserved a colour connection exists between the initial particles and the stable hadrons. Measuring the colour flow is crucial to validate the phenomenological description of this process. The jet-pull vector, ${\cal{\vec{P}}} (j)$ is a $p_{\mathrm{T}}$-weighted radial moment of jet $j$ and can be used to measure the colour flow~\cite{theorycolourFlow}. Using the jet-pull vector, also the jet-pull angle between any two jets can be calculated. The jet-pull angle, $\theta(j_1,j_2)$, is the angle between the jet-pull vector ${\cal{\vec{P}}} (j_1)$ and the vector connecting $j_1$ and $j_2$. The distribution of the jet-pull angle and the magnitude of the jet-pull vector have been measured in semileptonically decaying $t\bar{t}$ events created in proton collisions at 13 TeV~\cite{ATLAScolourFlow}. Some of these distributions are shown in Figure~\ref{fig:colourflow} for the data and different event generators. 
As expected, a clear colour flow is observed for the jets from the hadronically decaying $W$ boson, while the jet-pull angle distribution is nearly flat for the two jets from the $b$ quarks for which no colour connection is expected. None of the phenomenological descriptions is doing a good job for all distributions. The data for the colour flow between the two jets from the hadronically decaying $W$ boson is best described by \textsc{Powheg} interfaced with \textsc{Herwig}~7, while this model does the worst job of all generators for the colour flow between the two $b$ quark jets. 

\section{Charge asymmetry in top quark pair events}
\begin{figure}[htbp]
\centering
\includegraphics[width=0.55\textwidth]{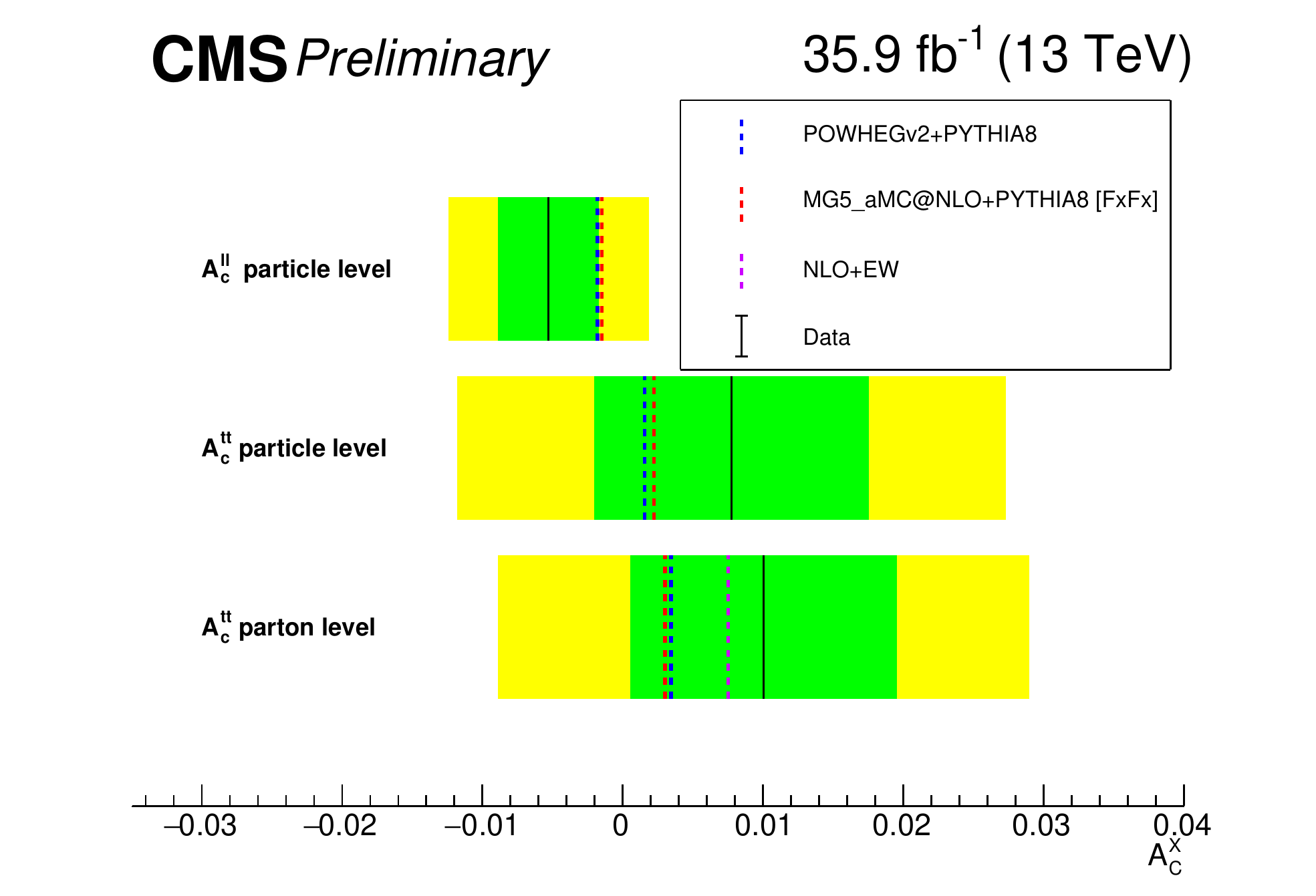}
\caption{The charge asymmetry measurements obtained from the reconstructed top quark and antiquark at parton and particle level or from the charged leptons (at particle level only)~\cite{TOP-17-014}. The dashed lines indicate the SM predictions produced with the \textsc{MGaMC$@$NLO} and \textsc{Powheg} generators both interfaced with \textsc{Pythia}~8, and the NLO+EW prediction from the SM.}
\label{fig:chargeasymm}
\end{figure}

Because of the boost of the initial (anti)quark the top (anti)quark is emitted in the direction of the incoming (anti)quark. The asymmetry is not present for the leading-order (LO) diagram or when the top quark pair is produced through gluon fusion. It is only present due to interference of the LO and box diagram on the one hand and the interference of the diagrams with initial and final state radiation on the other hand. Theories beyond the standard model could result in an enhanced asymmetry when a new boson is exchanged in the production.
At the LHC the direction of the (anti)quark is not known, but quarks have on average a larger momentum than antiquarks, since the latter can only be resolved from the sea. This means that top quarks will have on average larger rapidity values compared to top antiquarks. The charge asymmetry is defined as:
\begin{equation}
A_C = \frac{N^{\Delta |y|>0}-N^{\Delta |y|<0}}{N^{\Delta |y|>0}+N^{\Delta |y|<0}},
\end{equation}
where $N$ is a number of events and $\Delta |y|$ is the difference in absolute rapidity of the top quark and antiquark (or, alternatively, the positively and negatively charged lepton). The number of events $N^{\Delta |y|>0}$ and $N^{\Delta |y|<0}$ can be obtained from the normalized parton- and particle-level differential cross section measurement of dileptonically decaying $t\bar{t}$ events. The measurements performed by the ATLAS and CMS Collaborations during the LHC Run~1, have been combined and the combination was used to constrain several theories beyond the SM~\cite{Run1ChargeAsymm}. 
Recently, a first measurement of the charge asymmetry was performed at 13 TeV~\cite{TOP-17-014}. The resulting measurement is shown in Figure~\ref{fig:chargeasymm} and compared to the next-to-leading order (NLO) prediction in QCD, including corrections arising from mixing between QCD and electroweak diagrams, and between QCD and quantum electrodynamics (QED) diagrams~\cite{theorySpinMatrix}. Good agreement is observed between the predictions and the measurements.

\section{Spin correlations in top quark pair events}
\begin{figure}[htbp]
\centering
\includegraphics[width=0.43\textwidth]{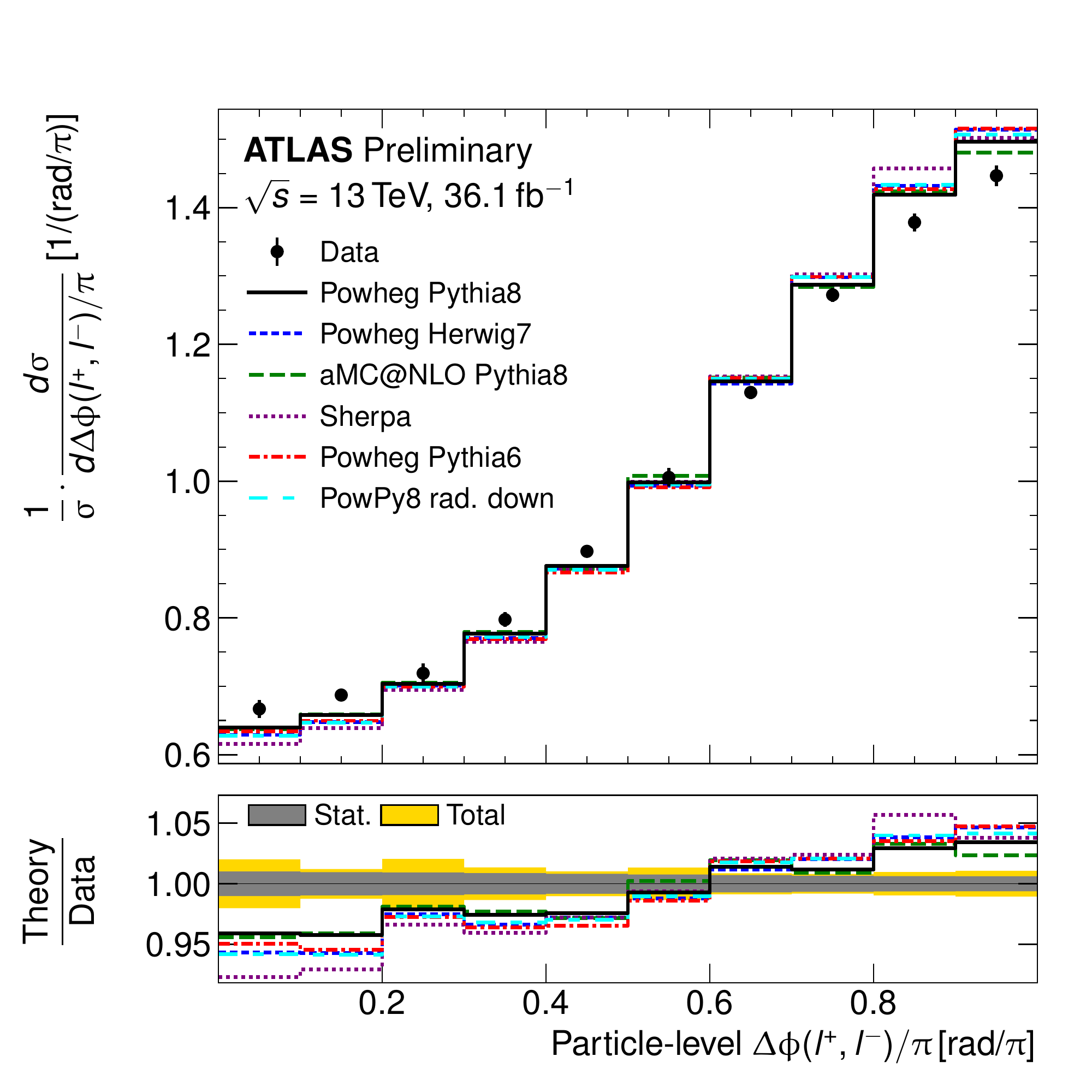}
\includegraphics[width=0.43\textwidth]{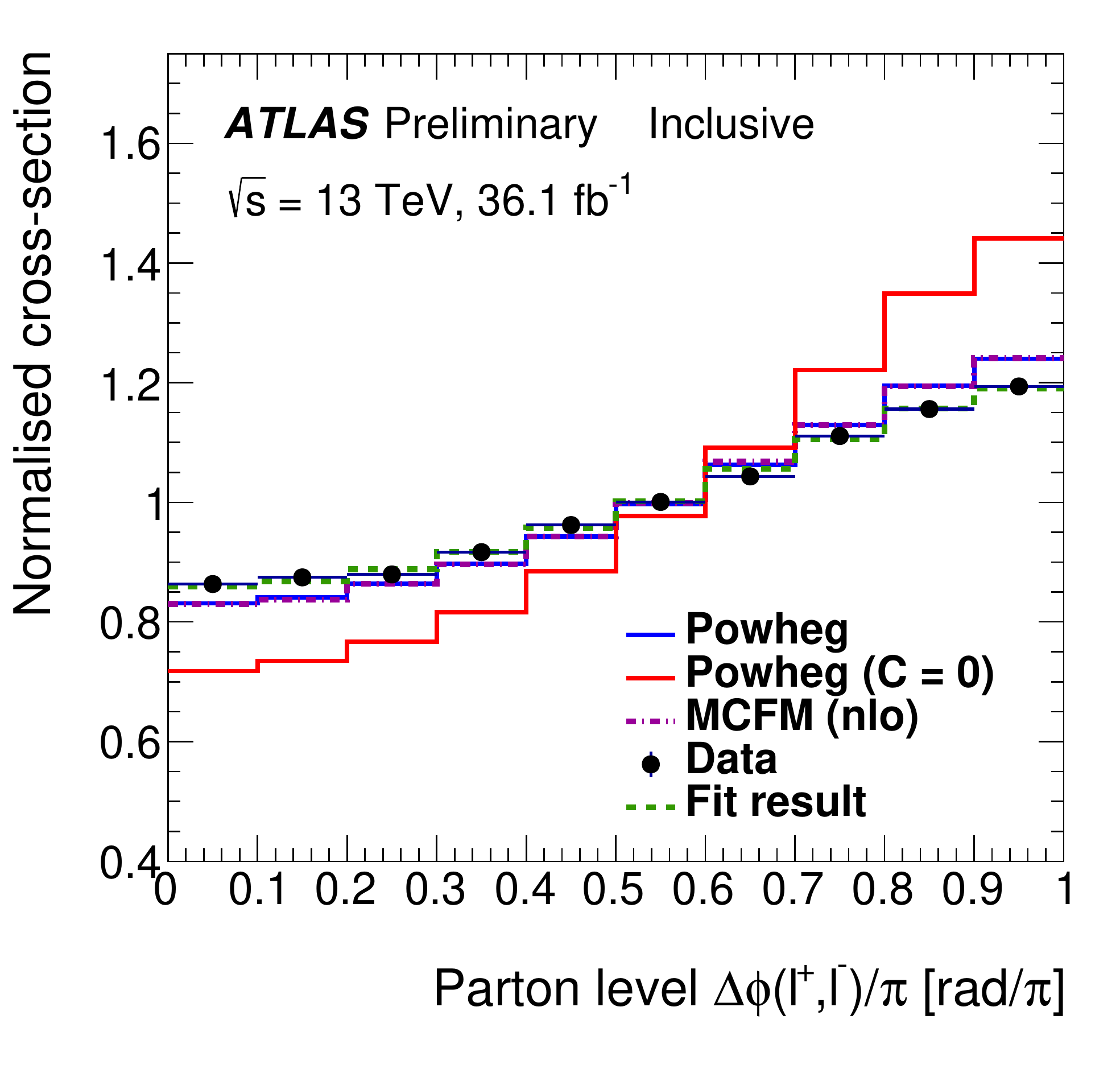}
\caption{The fiducial normalised particle-level differential cross-section compared to predictions from \textsc{Powheg}, \textsc{MGaMC$@$NLO} and \textsc{Sherpa} with different parton shower choices (left)~\cite{ATLAS-CONF-2018-027}. Result of the fit of hypothesis templates to the unfolded data (right)~\cite{ATLAS-CONF-2018-027}.}
\label{fig:spincorr}
\end{figure}

The spins of the top quark and antiquark are correlated with a strength depending on the spin quantization axis and the production process. The spin correlation can be measured via the angular distributions of decay products or using the matrix element method. A first spin correlation measurement has been performed at 13~TeV by measuring the distribution of the difference in azimuthal angle between the two charged leptons, $\Delta\phi(l^+,l^-)$ in dileptonically decaying $t\bar{t}$ events~\cite{ATLAS-CONF-2018-027}. This distribution is shown in Figure~\ref{fig:spincorr} and compared to a range of event generators with different parton shower choices. 
The data does not agree with the simulation and a fit yields a SM spin correlation fraction of $1.25\pm0.026\pm0.063$, which is a 3.2~$\sigma$ deviation from the expected unity value.
Since a deviation is observed, the spin correlation was also measured differentially in bins of $m_{t\bar{t}}$. This is particularly interesting since top quark pairs are mainly produced with (anti)parallel spins at low (high) $m_{t\bar{t}}$. However, no significant deviations are observed in bins of $m_{t\bar{t}}$ at this stage~\cite{ATLAS-CONF-2018-027}. During the workshop the point was raised that the mismodelling could potentially be resolved by including higher-order corrections in the decay~\cite{Poncelet}. 

\section{Summary}
Most of the LHC Run~1 and~2 top quark property measurements are limited by systematic uncertainties. From the experimental side these are typically the jet energy scale/resolution uncertainties. From the modeling side, the uncertainties due to the radiation, the event generator and the parton shower dominate. Most of the Run~2 measurements use only 25\% of the available proton collision data. Using more data will help to understand and further reduce the systematic uncertainties. Differential cross section measurements can for instance be used to further constrain the modeling uncertainties and a lot of progress is made to tune the event generators to the data. On the other hand, novel techniques can be exploited to trade statistical precision for a reduced systematic uncertainty. An example of such a technique is presented in~\cite{ReSyst}.


\end{document}